# Exploring 5G Network Performance: Comparison of Inner and Outer City Areas in Phetchaburi Province

**Phisit Pornpongtechavanich[1], Therdpong Daengsi[2]**
[1]Department of Information Technology, Faculty of Industry and Technology, Rajamangala University of Technology Rattanakosin Wang Klai Kangwon Campus, Prachuap Khiri Khan, Thailand
[2]Department of Sustainable Industrial Management Engineering, Faculty of Engineering, Rajamangala University of Technology Phra Nakhon, Bangkok, Thailand

| Article Info | ABSTRACT |
|---|---|
| *Article history:*<br><br>Received month dd, yyyy<br>Revised month dd, yyyy<br>Accepted month dd, yyyy<br><br>*Keywords:*<br><br>Quality of Service<br>Digital tourism<br>Download<br>Upload<br>Field test | The advancement of 5G technology has transformed various aspects of life, including tourism, by enabling people worldwide to communicate and travel with ease. Traveling to different places and countries is now seamless, removing language barriers and facilitating easy access to information on culture, accommodation, and tourist attractions. Additionally, access to applications that facilitate quicker language translation further enhances the travel experience. Phetchaburi Province holds significant importance as a global tourist destination. UNESCO has recognized Phetchaburi as a member of the UNESCO Creative Cities Network (UCCN), comprising one of 49 cities worldwide acknowledged for their creative city initiatives. Phetchaburi Province stands as the 5th city in Thailand to receive this designation. This research investigated 5G performance in Phetchaburi Province, both the inner and outer city, focusing on download and upload speeds. The results indicate that there is widespread 5G coverage throughout Phetchaburi Province, including urban and rural areas, especially for the 5G network with a good performance provided by one of the mobile network operators. In addition, the statistical analysis reveals differences in 5G performances between the inner city and the outer city of Phetchaburi Province, particularly for download speeds (p-value < 0.001).<br><br> |

*Corresponding Author:*

Therdpong Daengsi
Department of Sustainable Industrial Management Engineering, Faculty of Engineering, Rajamangala University of Technology Phra Nakhon, Bangkok, Thailand
1381 Pracharat Road 1, Wong Sawang Subdistrict, Bang Sue District, Bangkok 10800
Email: therdpong.d@rmupt.ac.th

## 1. INTRODUCTION

The data on the usage of 5G networks in Thailand has increased, as reported by the statistical data from the Office of the National Broadcasting and Telecommunications Commission. The National Telecommunications Commission (NBTC) stated in its latest report in May 2023 that there was a total of 52,160,445 mobile internet users, with a domestic bandwidth of 13,996.605 Gbps and an international bandwidth of 22,321.252 Gbps [1]. The high demand for 5G services in Thailand underscores their necessity for both the general public and businesses across various industries. With the advent of 5G mobile communications, Thailand has transitioned towards leveraging technology to foster national development. This includes its application in diverse sectors such as tourism, agriculture, industrial operations, education, healthcare, and more [2]. Additionally, the integration of 5G applications into homes has facilitated the concept of smart homes, enabling functionalities like remote door control, management of household appliances, lighting, security systems, and other innovations [3]. In the concept of smart cities, a network of interconnected traffic lights and sensors, for instance, detects vehicle movements, surpassing traditional time-based systems. Moreover, applications extend to surveillance through closed-circuit cameras (CCTV), facial analysis systems, car license plate reading systems, and even industrial applications like the Industrial Internet of Things (IIoT). The 5G system consists of three main components: 1) the 5G access network (5G Access Network: 5G-AN), 2) the core network (5G Core Network: 5GC), and 3) User Equipment (UE) [4]. Each component is deemed crucial to the system's operation. 5G is a rapidly developing technology that supports a wide range of wireless applications and makes use of a single-band antenna tuned to a particular resonance frequency [5-6].

In Thailand, the performance of 5G networks has been studied using field tests at BTS Skytrain stations along the Sukhumvit Line. The findings revealed an average download speed of 240.3 Mbps, an average upload speed of 87.3 Mbps, a latency of 19 ms, a jitter time of 8 ms, and a loss of 0.299 percent, respectively. It has been proposed to expand the scope of the study to include provincial areas to gain insights into the promotion and development of 5G networks by service providers in Thailand [7]. Factors influencing the use of network services encompass quality, stability, high speed, and coverage of network service areas [8]. Enhancing the efficiency of 5G network coverage in provincial areas is imperative for furthering economic expansion across various provinces.

Phetchaburi Province holds significance as a prominent tourist destination, recognized by the Educational, Scientific, and Cultural Organization of Thailand and designated by the United Nations Educational, Scientific, and Cultural Organization (UNESCO) as a City of Gastronomy in 2021, part of the UNESCO Creative Cities Network (UCCN). It joins 48 other cities worldwide in this distinction. Phetchaburi Province marks the fifth Thai city to be recognized in the Creative Cities Network, following Phuket (2015), Chiang Mai (2017), Sukhothai, and Bangkok (2019) [9]. The influx of tourists to Phetchaburi province remains substantial, comprising both Thai and foreign visitors. In 2022, Thai tourists numbered 10,670,442, marking a 19.36% increase from the previous year, while foreign tourists totaled 149,024 representing a staggering 198.46% increase. Revenue generated from Thai tourists amounted to 31,075.36 million baht, indicating a 28.48% rise from 2021, while revenue from foreign tourists reached 1,258.99 million baht, marking a remarkable 506.92% increase from the previous year [10]. This percentage represents a significant increase compared to the previous year, especially concerning the influx of foreigners visiting Phetchaburi province.

Therefore, advanced technology is imperative for elevating the level of tourism in Phetchaburi province, to make it more modern and accommodating for tourists from diverse linguistic and cultural backgrounds. Particularly, the 5G internet signal is another crucial factor in accessing information about key tourist attractions and facilitating communication during travel. Additionally, it is utilized for language translation to interact with the local community. For this reason, the research has focused on a comparative assessment of the efficiency of 5G networks in terms of downloading and uploading data within the inner city and outer city of Phetchaburi province. This study is significant as in-depth analyses of 5G Performance have not been conducted in this province before. The study employed three applications to assess the quality of 5G provided by two major mobile network operators (MNOs). The findings from this study reveal the actual download and upload speeds provided by these operators in Phetchaburi province. The methods and tools applied in this study can be used for further research in other provinces in Thailand and other cities worldwide, which is one of the contributions of this study.

This article has the following structure: After an Introduction section, Section 2 presents the background information, including literature reviews. Then, Sections 3 and 4 present the methodology and the results and analysis, respectively. Finally, Section 5 presents the discussion and conclusion.

## 2. BACKGROUND
### 2.1. Overview on 5G

The 5th generation mobile network, or 5G for short, is a system that has the ability to support communications and services, including text, sound, images, and animations [11], as well as supporting the use of broadband internet for users to access services and various applications, including multimedia commnucations (e.g., IP telephony, and online gaming [12-13]), along with being able to process large amount of data with only a small amount of processing time [14]. Because of this, it is crucial for businesses to compete with one another in offering a high quality of service (QoS) in order to satisfy customer needs. In Thailand, there are two main providers of 5G network services. Both companies use 2600 MHz 5G spectrum and a few bands to deliver commercial services while adhering to 5G network standards [7][15]. Regulations have been established. These can be summarized as shown in Figure 1. Along with meeting the needs of the majority of people, these 5G specifications will also support applications in various industries, [16] such as 1-10Gbps connections to the endpoint, round-trip delays of 1ms end-to-end latency, 1,000x bandwidth per unit area, 10-100x number of connected devices, 99.999% perceived availability, perceived coverage of 100%, a 90% reduction in network energy consumption, and up to ten years of battery life for low-power machine-type devices [17], etc.

## 2.2. Literature review

Many researchers have conducted extensive research on 5G, focusing on its efficiency, quality, service provision, and other aspects of the technology in their respective countries. All research is centered on the efficiency of 5G as a wireless network, which can be summarized as shown in Table 1 [7], [18-33]. One can see that many research studies have focused on 5G performance. However, most were conducted in Western countries. Therefore, there is a rationale for analyzing the data obtained from the field tests in Phetchaburi Province, Thailand.

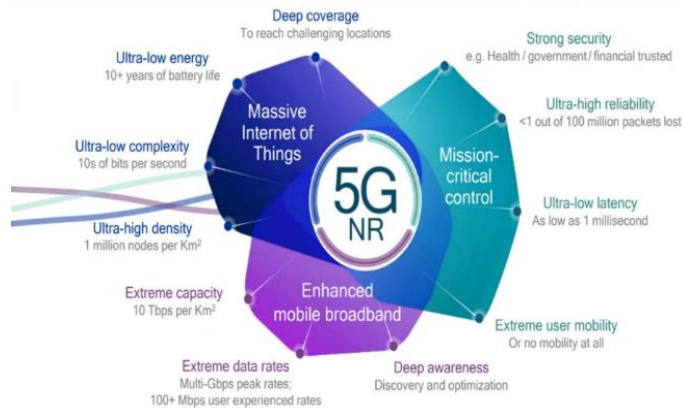

Figure 1. 5G Network Requirements [16]

Table 1. Related Works

| Ref. | Study of research | Network | | | Tools | Country |
|---|---|---|---|---|---|---|
| | | 4G | 5G | 6G | | |
| [7] | Analizing 5G performance (using data from Opensignal application) at 60 BTS Skytrain stations in Bangkok. | | ✓ | | Mobile Phones | Thailand |
| [18] | An analysis comparing 4G and 5G network design, performance, frequency spectrum bands, multiplexing methods, and applications. | ✓ | ✓ | | Mobile Phones | Romania |
| [19] | Enhancing the proactive policy effectiveness of 5G networks in the EU. | | ✓ | ✓ | Data Analysis | Italy |
| [20] | Techno-economic comparison of cost models for software-protected networks (Cognitive Radio) and software-defined networks (SDN) in 5G networks. | | ✓ | | Create a Model | Greece |
| [21] | Comparing information on beamforming, latency, MIMO, bandwidths, and signal strengths. | ✓ | ✓ | | Mobile Phones | Ecuador |
| [22] | Performance Comparison of Private and Public Blockchains of 5G Networks. | | ✓ | | Signal Comparison | Russia |
| [23] | To compare the industrial IoT's data download performance. | | ✓ | ✓ | Mobile Phones | Finland |
| [24] | Comparison between 5G indoor signals in the 3.6-GHz and 26-GHz bands. | | ✓ | | Simulations System | Finland |
| [25] | Develop an improved AODV protocol solution for MANETs used in 5G to be utilized as a routing algorithm to find 5G QoS paths. | | ✓ | | Develop solutions and algorithms | Vietnam |
| [26] | Predicting the demand for indoor 5G connectivity in locations such as university campuses. | | ✓ | | Forecast | Finland |
| [27] | Information comparison: RSRP, RSRQ, and SINR depending on signal coverage in 4G LTE Network and 5G New Radio regions. | ✓ | ✓ | | Mobile Phones | Nigeria |
| [28] | Performance testing of the Downlink throughput, Uplink throughput, Latency, and Packet loss of 5G wireless networks. | | ✓ | | Lab test simulation | Sweden |
| [39] | 5G performance evaluation by comparing the 5G waveforms including OFDM, UFMC, FBMC, and GFDM. | | ✓ | | specified situations/tools | France |
| [30] | The study examined the latency and throughput performance of 4G and 5G signals in Chicago and Miami in a variety of conditions, including outdoor, indoor-outdoor, and extreme temperatures. | ✓ | ✓ | | Mobile Phone | USA |
| [31] | Review of 5G network issues, such as the open issues, research challenges, channel estimation, multi-carrier modulation, and 5G applications. | | ✓ | | Literature Review | India |
| [32] | Comparative study of 5G speeds in the heart of Phetchaburi province Thailand. | | ✓ | | Mobile Phones | Thailand |
| [33] | Performance measurement of 4G and 5G networks, LTE, NB-IoT, and 5G NR connectivity. | ✓ | ✓ | | Mobile Phones | Italy |

## 3. METHODOLOGY

In order to assess the performance of 5G networks in Phetchaburi province provided by two operators, hereafter called Oper-X and Oper-Y, which are the major MNOs in Thailand that cover more than 90 million mobile subscribers in total, the research methods, , in this study were divided into three phases, as shown in Figure 2.

Phase 1: involves studying, analyzing, and synthesizing documents and research at the national and international levels related to the performance of 4G, 5G, and 6G networks between 2018 and 2024. This phase aims to summarize the findings, analyze data from 34 research papers as presented in the Introduction and Background sections, and subsequently design the research process for Phase 2.

Phase 2: involves designing a data collection methodology utilizing a portable mobile device, specifically the Vivo V23e 5G. It is equipped with a chipset capable of connecting to 5G Dual Mode (SA/NSA) simultaneously on both SIM cards (Dual 5G SIM), and a graphics processing unit (GPU) Mali-G57 MC2. It operates with 8GB of RAM and runs on the Android operating system. It is powered by 12 chipsets and an Octa-Core processor. This mobile phone is the same phone as mentioned in [7]. This data collection design aimed to gather the data, consisting of download and upload speeds from 5G networks provided by Oper-X and Oper-Y. The data gathering was conducted at various significant places or locations, covering the inner city of Phetchaburi and the outer city in the late Q1/2023. The inner city (See Figure 3) consisted of 11 important places: Wat Kamphaeng Laeng, Wat Yai Suwannaram, Wang Ban Puen, Tham Khao Luang, Wat Khoi, Sanam Luang Phetchaburi, Wat Tumklaeb, Khao Wang, Wat Pranon, Wat Khao Ban Dai It, and Wat Mahathat Worawihan. The outer city (see Figure 3), consisted of 14 important places, including Wat Nerancharararam, Puek Tian Beach, Wat Nai Pak Ta le, Wat Khao Takrao, Khao Nang Panthurat, Chao Samran Beach, Wat Nok Pak Ta Le, Cha-am Beach, Wat Phet Suwan, Pu Chak-Saphan Yok, Nana Sirindhorn, Cha-am Forest Park, Naresuan Camp, and Wat Nai Klang. Three mobile applications were used to test speed, consisting of nPerf, Opensignal, and Speedtest, as in [7]. Data collection was conducted using these three mobile applications, with field tests at many significant tourist attractions in Phetchaburi province. These applications were utilized to evaluate 5G performance at three test points per place with two rounds (one in the morning and one in the afternoon). Each field test at each place was randomly conducted between 08:00 am and 06:00 pm.

Phase 3: This step involved data collection from Phase 2, which comprised a total of 900 items (450 items from Oper-X and 450 items from Oper-Y). The data were processed and calculated for average download and upload speeds, before conducting the statistical analysis, as shown in Figure 2. The results and the analysis are presented in the next section.

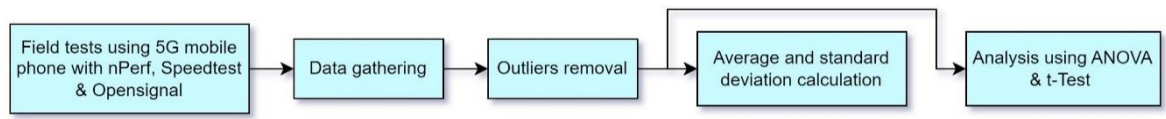

Figure 2. Overview of the processes in this study

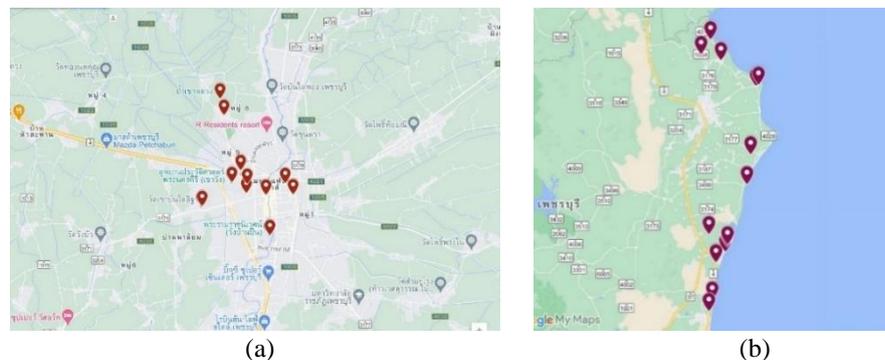

(a) (b)

Figure 3. Important tourist attractions in Phetchaburi (a) within the inner city and (b) within the outer city

## 4. RESULTS AND ANALYSIS
### 4.1. Results of overall speeds

As shown in Figure 4(a), the results depict the collected 5G data in terms of downloading. Data were collected from a total of 25 locations, comprising one morning round and one afternoon round at each location. The testing duration was between 08:00 and 18:00. The data analysis showed that the Speedtest on the Oper-X network exhibited an average download speed of 175.6 Mbps, while on the Oper-Y network, it was 49.1 Mbps. For the nPerf application, the Oper-X network showed an average download speed of 146.9 Mbps, compared to 48.1 Mbps on the Oper-Y network. Lastly, the Opensignal application recorded an average download speed of 153.6 Mbps on the Oper-X network and 46.6 Mbps on the Oper-Y network. Figure 4(b) presents the results of the 5G data collected in terms of uploading. The data analysis showed that the Speedtest on the Oper-X network exhibited an average upload speed of 31.7 Mbps, while on the Oper-Y network, it was 16.2 Mbps. The nPerf applicationusing the Oper-X network showed an average upload speed of 23.0 Mbps, compared to 13.4 Mbps on the Oper-Y network. Lastly, the Opensignal application recorded an average upload speed of 27.4 Mbps on the Oper-X network and 17.0 Mbps on the Oper-Y network.

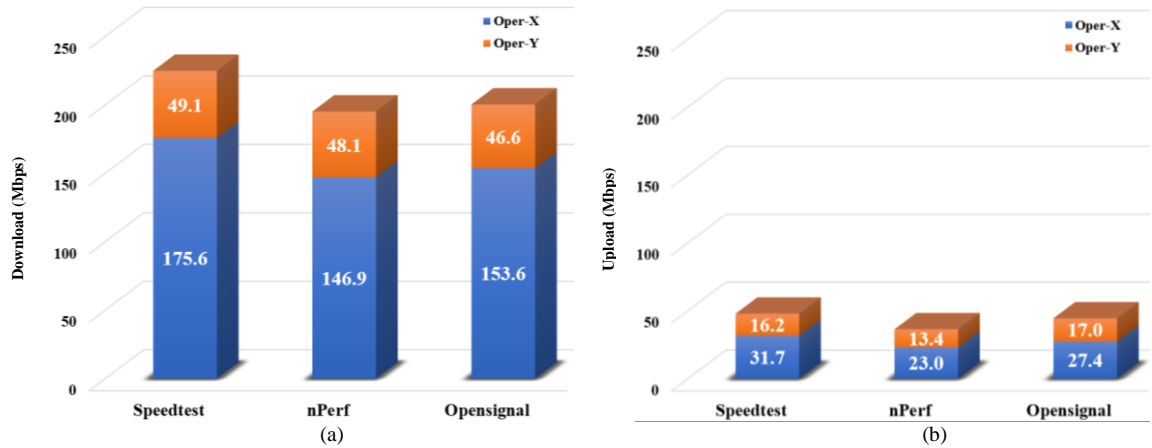

Figure 4. 5G speeds obtained from three applications in Phetchaburi (a) Download speeds (b) Upload speeds

**4.2. Results of 5G data speeds at specific locations within the inner city of Phetchaburi**

As presented in Table 2, the results depict the downloading and uploading data of 5G networks within the inner city of Phetchaburi, encompassing a total of 11 locations. The analysis of the data for each location, as presented in the table, provides information on the average values obtained from collecting data from all three applications and averaging them for each location. The data reveal that within the inner city of Phetchaburi, the Oper-X network exhibited an average download speed of $196.4 \pm 67.8$ Mbps and an upload speed of $26.6 \pm 11.0$ Mbps. Conversely, the Oper-Y network recorded an average download speed of $68.7 \pm 47.9$ Mbps and an upload speed of $16.0 \pm 11.4$ Mbps. The average from both MNOs, the average download and upload speeds are $132.6 \pm 57.8$ Mbps and $21.3 \pm 11.2$ Mbps, respectively. Notably, Wat Khoi demonstrated the highest download speed, followed by Sanam Luang Phetchaburi. Sanam Luang Phetchaburi showed the highest upload speed, followed by Wat Khoi.

**4.3. Results of 5G data speeds at specific locations within the outer city of Phetchaburi**

As presented in Table 3, the results of the download and upload data speeds of 5G networks outside the inner city of Phetchaburi, encompassing a total of 14 locations, are presented. The data has been analyzed separately for each location, as depicted in the table, providing insights into the performance in each area. It is evident from the information that in the outer city of Phetchaburi, across the 14 locations, the Oper-X network recorded an average download speed of $129.1 \pm 72.6$ Mbps and an upload speed of $28.0 \pm 14.4$ Mbps. While the average download speed for the Oper-Y network, was $31.5 \pm 21.1$ Mbps, with an upload speed of $15.1 \pm 8.6$ Mbps. The averages for both operators were download and upload speeds of $80.3 \pm 46.8$ Mbps and $21.5 \pm 11.5$ Mbps, respectively. In addition, Wat Nai Klang emerged as the location with the highest download speed, followed by Naresuan Camp, respectively. Chao Samran Beach exhibited the highest upload speed, followed by Pu Chak-Saphan Yok, as illustrated in the Table 3.

**4.4. Results of the Analysis**

After obtaining the data on the download and upload speeds provided by 5G network operators from the test files within both the inner and outer areas of Phetchaburi, a statistical analysis using the t-Test technique was conducted on the basis of eight hypotheses. The results of the analysis are presented in Table 4.

H1: The download speeds via 5G provided by Oper-X and Oper-Y within the inner city are the same or not.
H2: The download speeds via 5G provided by Oper-X and Oper-Y within the outer city are the same or not.
H3: The upload speeds via 5G provided by Oper-X and Oper-Y within the inner city are the same or not.
H4: The upload speeds via 5G provided by Oper-X and Oper-Y within the outer city are the same or not.
H5: The download speeds via 5G within the inner city and the outer city provided by Oper-X are the same or not.
H6: The download speeds via 5G within the inner city and the outer city provided by Oper-Y are the same or not.
H7: The upload speeds via 5G within the inner city and the outer city provided by Oper-X are the same or not.
H8: The upload speeds via 5G within the inner city and the outer city provided by Oper-Y are the same or not.

Table 4 is an analysis of the data utilizing a t-Test statistical technique to analyze the data speeds provided by 5G networks regarding downloading and uploading within the inner and outer city of Phetchaburi. The analysis of the hypotheses in each aspect can be summarized as follows:

- H1 and H2 show that the p-values are less than 0.001. This means that the download speeds within the inner and outer city of Phetchaburi provided by Oper-X network are significantly better than Oper-Y network.
- H3 and H4 show that the p-values are less than 0.001, which is consistent with the results for H1 and H2. This means that the upload speeds within the inner and outer city of Phetchaburi provided by Oper-X network are significantly higher than those of the Oper-Y network.
- H5 and H6 show that the p-values are less than 0.001. This means that the download speeds within the inner city of Phetchaburi provided by each MNO are significantly better than those in the outer city.

Table 2. Results of data download and upload via 5G networks within the inner city of Phetchaburi

| Place | Download (Mbps) | | Upload (Mbps) | |
| --- | --- | --- | --- | --- |
| | Oper-X | Oper-Y | Oper-X | Oper-Y |
| Wat Kamphaeng Laeng | 234.8 ± 58.91 | 161.6 ± 103.72 | 25.7 ± 7.6 | 18.2 ± 10.3 |
| Wat Yai Suwannaram | 180.4 ± 56.55 | 45.2 ± 18.87 | 35.2 ± 10.8 | 22.7 ± 12.7 |
| Wat Pranon | 160.1 ± 63.45 | 141.3 ± 83.35 | 28.5 ± 15.0 | 26.1 ± 12.1 |
| Wat Khoi | 351.4 ± 60.75 | 63.2 ± 13.19 | 58.4 ± 28.2 | 16.2 ± 6.2 |
| Wat Khao Ban Dai It | 158.7 ± 140.33 | 26.2 ± 21.65 | 18.6 ± 11.3 | 14.6 ± 12.0 |
| Wat Mahathat Worawihan | 266.5 ± 89.09 | 25.0 ± 9.08 | 29.5 ± 14.5 | 7.2 ± 6.7 |
| Wang Ban Puen | 281.5 ± 58.21 | 30.0 ± 12.08 | 26.4 ± 9.0 | 11.5 ± 10.7 |
| Tham Khao Luang | 19.0 ± 7.27 | 14.0 ± 14.23 | 6.6 ± 6.4 | 3.0 ± 3.8 |
| Wat Tumklaeb | 121.8 ± 80.40 | 28.1 ± 13.94 | 5.3 ± 3.9 | 14.8 ± 10.9 |
| Sanam Luang Phetchaburi | 245.3 ± 60.22 | 155.7 ± 146.59 | 49.4 ± 9.0 | 34.5 ± 29.9 |
| Khao Wang | 140.7 ± 70.74 | 65.7 ± 90.03 | 9.1 ± 5.8 | 7.7 ± 9.7 |
| Average | 196.4 ± 67.8 | 68.7 ± 47.9 | 26.6 ± 11.0 | 16.0 ± 11.4 |
| | 132.6 ± 57.8 | | 21.3 ± 11.2 | |

Table 3. Results of data download and upload via 5G networks within the outer city of Phetchaburi

| Place | Download (Mbps) | | Upload (Mbps) | |
| --- | --- | --- | --- | --- |
| | Oper-X | Oper-Y | Oper-X | Oper-Y |
| Chao Samran Beach | 81.0 ± 41.4 | 31.2 ± 21.6 | 49.1 ± 21.6 | 19.5 ± 8.5 |
| Puek Tian Beach | 48.1 ± 26.2 | 8.6 ± 5.4 | 25.3 ± 15.1 | 6.2 ± 3.2 |
| Cha-am Beach | 120.0 ± 120.1 | 32.2 ± 26.7 | 29.7 ± 18.2 | 20.6 ± 5.5 |
| Wat Nerancharraram | 20.0 ± 11.9 | 15.9 ± 9.8 | 18.7 ± 9.9 | 12.9 ± 6.8 |
| Pu Chak-Saphan Yok | 159.1 ± 173.6 | 24.8 ± 9.3 | 43.4 ± 19.8 | 21.2 ± 7.9 |
| Khao Nang Panthurat | 88.5 ± 71.0 | 22.1 ± 11.7 | 9.9 ± 6.9 | 10.4 ± 5.0 |
| Cha-am Forest Park | 198.9 ± 88.1 | 19.5 ± 10.2 | 27.5 ± 18.3 | 8.6 ± 8.5 |
| Nana Sirindhorn | 179.8 ± 104.2 | 20.3 ± 12.8 | 29.1 ± 16.7 | 11.3 ± 11.5 |
| Naresuan Camp | 207.0 ± 68.1 | 65.7 ± 62.5 | 24.0 ± 14.3 | 14.6 ± 14.8 |
| Wat Khao Takrao | 102.1 ± 30.7 | 7.0 ± 4.3 | 15.8 ± 6.8 | 6.2 ± 2.7 |
| Wat Nai Klang | 321.0 ± 7.3 | 73.5 ± 65.5 | 37.1 ± 11.8 | 15.6 ± 11.0 |
| Wat Phet Suwan | 127.5 ± 130.3 | 30.5 ± 23.6 | 32.1 ± 14.5 | 18.3 ± 15.0 |
| Wat Nok Pak Ta Le | 79.6 ± 43.1 | 67.7 ± 24.3 | 21.2 ± 15.9 | 29.6 ± 16.6 |
| Wat Nai Pak Ta le | 74.7 ± 37.0 | 22.70 ± 7.5 | 28.6 ± 11.7 | 16.5 ± 3.2 |
| Average | 129.1 ± 72.6 | 31.5 ± 21.1 | 28.0 ± 14.4 | 15.1 ± 8.6 |
| | 80.3 ± 46.82 | | 21.53 ± 11.49 | |

Table 4. Analyzed results

| Hypothesis | t-Test | p-value | Results |
| --- | --- | --- | --- |
| H1 | 7.486 | <0.001 | Significant |
| H2 | 7.849 | <0.001 | Significant |
| H3 | 3.611 | <0.001 | Significant |
| H4 | 6.508 | <0.001 | Significant |
| H5 | 3.731 | <0.001 | Significant |
| H6 | 3.389 | <0.001 | Significant |
| H7 | 0.489 | 0.626 | Insignificant |
| H8 | 0.453 | 0.651 | Insignificant |

Remark: p-value < 0.05 means significant with 95% confidence interval

- However, H7 and H8 show that the p-values are 0.626 and 0.651, respectively. This means that there is no significant difference between the upload speeds within the inner and outer city of Phetchaburi provided by each of the mobile network operators.

## 5. DISCUSSION

After conducting this study and analyzing the results, several issues need to be considered:

- This study is similar to [32], as the data utilized were from the same source. However, this study advances beyond [33] by covering not only the field tests within the inner city of Phetchaburi, but also those within the outer city, including several important tourist attractions. Furthermore, this study includes results and analysis from not only the nPerf application, but also the Speedtest and Opensignal applications.

- When compared to the average download and upload speeds of 132.6 Mbps and 21.3 Mbps, respectively, from the inner city, and the average download and upload speeds of 80.3 Mbps and 21.5 Mbps, respectively, from the outer city of Phetchaburi, these speeds are lower than the average download and upload speeds of 140.4 Mbps and 52.0 Mbps, respectively, from field tests in Bangkok as reported in [7].
- Comparing the overall average download speed of 132.6 Mbps from the inner city field tests to the overall average download speed of 80.3 Mbps from the outer city field tests (as shown in Tables 2-3), it is evident that the speeds from the inner city tests were higher than those from the outer city tests. This suggests that the mobile network operators provide a higher quality of service to users within the inner city.
- However, when comparing the overall average upload speed of 21.3 Mbps from the inner city tests to the overall average upload speed of 21.5 Mbps from the outer city tests (as shown in Tables 2-3), it is apparent that the speeds are almost the same. This indicates that mobile network operators provide a similar quality of service to users in both the inner and outer city.
- According to the statistical analysis of the hypotheses shown in Table 4, the results confirm significant differences in all aspects of download speeds (p-values < 0.05). However, there is no significant difference in upload speeds between the inner and outer city provided by the same mobile network operator (p-values > 0.05).
- This paper may be the first to present an in-depth study and analysis of 5G performance within Phetchaburi Province. Therefore, the information revealed in this paper should be considered by operators to improve their 5G networks and services in the future, particularly in the rural areas outside the city.

## 6. CONCLUSION

This study shows the status of 5G coverage in a city in Thailand. Data gathered in 2023 showed that, within the inner city of Phetchaburi, the 5G mobile network provided by Oper-X had average download and upload speeds of 196.4 Mbps and 26.6 Mbps, respectively, while the 5G network provided by Oper-Y had average download and upload speeds of 68.7 Mbps and 16.0 Mbps, respectively. Furthermore, outside the inner city of Phetchaburi, the 5G mobile network provided by Oper-X had average download and upload speeds of 129.1 Mbps and 28.0 Mbps, respectively, whereas the 5G network provided by Oper-Y had average download and upload speeds of 31.5 Mbps and 15.1 Mbps, respectively. This study demonstrates that Phetchaburi Province, one of the UNESCO Creative Cities, has sufficient 5G area coverage, as 5G signals provided by the two major 5G mobile network operators cover all important tourist attractions and locations. This means that internet connectivity is available and meets the needs of both visitors, including foreigners and local users in this province, especially at key tourist sites within and outside the inner city of Phetchaburi.

Despite the differing performance of the two 5G mobile networks, the data analysis in this research provides valuable insights for considering future signal improvements by the mobile network operators, especially Oper-Y, in Phetchaburi Province. Therefore, Oper-Y should consider the results from this study for their network improvement. Meanwhile, the NBTC—the regulator—may verify whether the performance of the 5G network meets the required service levels. Last but not least, since this study was conducted Phetchaburi, a province in southern Thailand, it can serve as an example for 5G field tests. Specifically, the methods, tools, and processes described in this paper can be utilized to study other cities to assess the performance of mobile networks in those areas. Nonetheless, this study did not cover other QoS metrics, including reference signal received power (RSRP), reference signal received quality (RSRQ), signal-to-noise ratio (SNR), latency, jitter, and packet loss. These should be considered for study in future work.

**CONFLICT OF INTEREST**

The authors declare that there is no conflict of interest.

**ACKNOWLEDGEMENTS**

Thanks to Rajamangala University of Technology Rattanakosin and Rajamangala University of Technology Phra Nakhon, for supporting this study. Thanks to Jutharat Inseenan and Assoc. Prof. Dr. Pongpisit Wuttidittachotti for collecting and sharing the data. Lastly, thanks to Peter Bint for English editing.

## BIOGRAPHIES OF AUTHORS

| | |
|---|---|
| 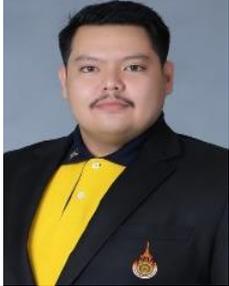 | **Phisit Pornpongtechavanich**  is an Assistant Professor in the Faculty of Industry and Technology, Rajamangala University of Technology Rattanakosin, Wang Klai Kangwon Campus (RMUTR_KKW). In 2012, he received his Bachelor of Technology degree in information technology from RMUTR_KKW. He obtained a scholarship and then received a Master of Science in information technology from KMUTNB in 2014 and a Ph.D. in Information and Communication Technology for Education in 2023. His research interests include security, Deep Learning, AI, IoT, VoIP quality measurement, QoE/QoS, Mobile Networks, and Multimedia Communications. He can be contacted at email: phisit.kha@rmutr.ac.th. (Note: he is also the co-corresponding author for this paper.) |
|  | **Therdpong Daengsi**  is an Assistant Professor in the Faculty of Engineering, RMUTP. He received a B.Eng. in Electrical Engineering from KMUTNB in 1997. He received a Mini-MBA Certificate in Business Management and M.Sc. in Information and Communication Technology from Assumption University in 2006 and 2008, respectively. Finally, he received Ph.D. in Information Technology from KMUTNB in 2012. He also obtained certificates including Avaya Certified Expert – IP Telephony and ISO27001. With 19 years of experience in the telecom business sector, he has also worked as an independent academic for a short period before being a full-time lecturer at present. His research interests include QoS/QoE, mobile networks, multimedia communication, telecommunications, cybersecurity, data science, and AI. He can be contacted by email: therdpong.d@rmutp.ac.th. |